\documentclass[journal]{IEEEtran}

\IEEEoverridecommandlockouts
\usepackage{cite}
\usepackage{amsmath,amssymb,amsfonts}
\usepackage{algorithmic}
\usepackage{graphicx}
\usepackage{textcomp}
\usepackage{xcolor}
\usepackage{dsfont}
\usepackage{multirow}
\usepackage{booktabs}
\usepackage{bm} 

\usepackage{caption} 
\usepackage{subcaption} 

\def\BibTeX{{\rm B\kern-.05em{\sc i\kern-.025em b}\kern-.08em
    T\kern-.1667em\lower.7ex\hbox{E}\kern-.125emX}}

\begin{document}

\title{BeamVLM for Low-altitude Economy: Generative Beam Prediction via Vision-language Models}

\author{Chenran~Kou,
Changsheng~You,
Mingjiang~Wu,
Dingzhu~Wen,
Zezhong~Zhang,
and~Chengwen~Xing

\thanks{Chenran~Kou, Changsheng~You, and Mingjiang~Wu are with the Department of Electronic and Electrical Engineering,
Southern University of Science and Technology, Shenzhen 518055, China
(e-mail: chenrankou63@gmail.com; \{youcs, wumj\}@sustech.edu.cn).}

\thanks{Dingzhu~Wen is with the Network Intelligence Center, School of Information Science and Technology, ShanghaiTech University, Shanghai, China
(e-mail: wendzh@shanghaitech.edu.cn).}

\thanks{Zezhong~Zhang is with the School of Science and Engineering (SSE), Shenzhen Future Network of Intelligence Institute (FNii-Shenzhen),
The Chinese University of Hong Kong (Shenzhen), Shenzhen 518172, China
(e-mail: zhangzezhong@cuhk.edu.cn).}

\thanks{Chengwen~Xing is with the School of Information and Electronics, Beijing Institute of Technology, Beijing 100081, China
(e-mail: xingchengwen@gmail.com).}
\thanks{\emph{Corresponding author: Changsheng You.}}
}

\maketitle
\begin{abstract}
For low-altitude economy (LAE), fast and accurate beam prediction between high-mobility unmanned aerial vehicles (UAVs) and ground base stations is of paramount importance, which ensures seamless coverage and reliable communications. However, existing deep learning-based beam prediction methods lack high-level semantic understanding of dynamic environments, resulting in poor generalization. On the other hand, the emerging large language model (LLM) based approaches show promise in enhancing generalization, but they typically lack rich environmental perception, thereby failing to capture fine-grained spatial semantics essential for precise beam alignment. To tackle these limitations, we propose in this correspondence a novel \emph{end-to-end generative} framework for beam prediction, called BeamVLM, which treats beam prediction as a \emph{vision question answering task} capitalizing on powerful existing vision-language models (VLMs). By projecting raw visual patches directly into the \emph{language domain} and judiciously designing an \emph{instructional prompt}, the proposed BeamVLM enables the VLM to jointly reason over UAV trajectories and environmental context. Last, experimental results on real-world datasets demonstrate that the proposed BeamVLM outperforms state-of-the-art methods in prediction accuracy and also exhibits superior generalization for other scenarios such as vehicle-to-infrastructure (V2I) beam prediction. 
\end{abstract}


\section{Introduction}
The rapid dynamics of low-altitude economy (LAE) pose significant challenges for unmanned aerial vehicle (UAV) beam management~\cite{ywj2025LAE}. Due to the high mobility and unpredictable orientation of UAVs, the channel coherence time is often very short for conventional beam training methods. This necessitates the use of beam prediction to maintain reliable links without prohibitively high overhead in exhaustive beam search.

To alleviate the overhead associated with beam prediction, various strategies have been proposed in the literature (see, e.g.,~\cite{youuav, jiang2022computer,Charan2022_UAV,camTVT}). 
Specifically, traditional non-vision-based approaches focus on optimizing beam search and beamforming procedures through hybrid offline--online optimization frameworks for UAV-enabled communications~\cite{youuav}. On the other hand, sensing-aided beam prediction has emerged as a promising alternative, leveraging multi-modal sensory data to assist beam alignment~\cite{jiang2022computer,Charan2022_UAV,camTVT}. For instance, base stations (BSs) can proactively infer the optimal beam directions for flying UAVs by utilizing aerial imagery from ground-based cameras~\cite{Charan2022_UAV}. To process such high-dimensional sensory data, existing research primarily relies on deep learning (DL) architectures, such as convolutional neural networks (CNNs) and recurrent neural networks (RNNs). While effective in specific scenarios, these methods often treat beam prediction as a black-box mapping task. Consequently, they tend to overfit to low-level pixel statistics and lack a high-level semantic understanding of scenes, resulting in limited generalization performance when deployed in unseen environments.

Recently, large language models (LLMs) have emerged as a promising paradigm for beam prediction in wireless communication systems. In particular, the vision for 6G networks emphasizes advanced transceiver technologies that integrate sensing, learning, and semantic-aware processing as native components of the physical layer and link adaptation mechanisms~\cite{YOU}. 
For example, the authors in \cite{zheng2025beamllmvisionempoweredmmwavebeam} proposed an LLM-based beam prediction method, which adopts a ``detection-then-reasoning'' pipeline based on object-level bounding boxes. Such a framework generally suffers from substantial modality information loss. Specifically, since it reduces a rich visual scene to isolated object proxies, it overlooks critical environmental context (e.g., building structures, urban canyon geometry, and potential blockers) that is inherently related to wireless channels. Consequently, these methods fail to capture fine-grained spatial correlations required for high-precision beam prediction.

Motivated by the above, we propose in this correspondence a novel end-to-end framework, called \emph{BeamVLM}, which treats beam prediction for UAVs as a \emph{generative} visual question answering (VQA) task by using the well-trained vision-language model (VLM). Unlike existing methods that rely on sparse text or bounding boxes, BeamVLM projects raw visual patches \emph{directly} into the language domain, hence allowing the model to jointly reason over high-dimensional visual semantics and task-specific instructions. Moreover, by judiciously designing an instructional prompt, our approach captures both the flight trajectories of the target UAV and the environmental context to perform consistent beam prediction. Finally, numerical results verify the effectiveness of the proposed BeamVLM framework in a dynamic UAV scenario as well as its generalization in the vehicle-to-infrastructure (V2I) scenario.

\section{System Model and Problem Formulation}
\subsection{System Model}
\label{ssec:system_model}
\begin{figure}[t]
    \centering
    \includegraphics[width=0.65\linewidth]{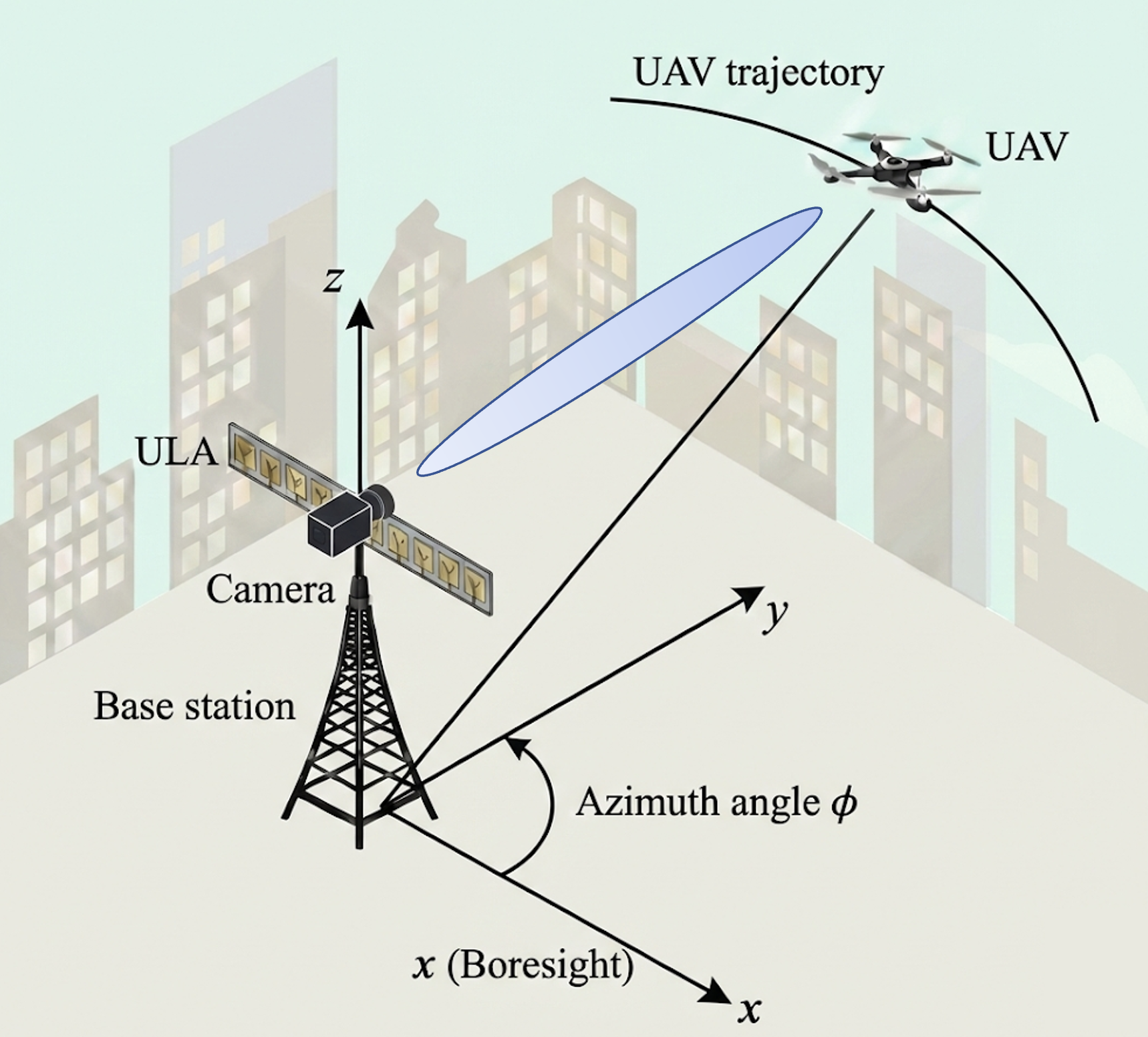} 
    \caption{Illustration of the ground-to-UAV communication system.}
    \label{fig:system_model}
\end{figure}

We consider a downlink ground-to-UAV millimeter-wave (mmWave) communication system as illustrated in Fig.~\ref{fig:system_model}, where a ground BS equipped with an $N$-antenna uniform linear array (ULA) communicates with a single-antenna UAV.\footnote{The model can be readily extended to multi-UAV scenarios by performing independent beam prediction for each UAV.} In addition to the antenna array, the BS is also equipped with an RGB camera that captures surrounding environment and the UAV's flight trajectory for sensing and prediction purposes. 

Let $\mathbf{h}[t] \in \mathbb{C}^{N \times 1}$ denote the channel from the BS to the UAV at time slot $t$. Based on publicly available real-world dataset in~\cite{DeepSense}, we consider the case where the BS applies beam prediction in the azimuth angle domain, denoted by $\phi$.\footnote{Extension to elevation prediction is straightforward using planar arrays.}
Moreover, accounting for the $90^\circ$ field-of-view of the deployed camera, a $90^\circ$ azimuth angular sector is considered for beam prediction, i.e., $\phi \in [-45^\circ, 45^\circ]$~\cite{DeepSense}.
To provide finer angular resolution in the beam prediction, we consider an oversampled discrete Fourier transform (DFT) beamforming codebook in the considered angular regime which is denoted by $\mathcal{F}=\{\mathbf{f}_1,\dots,\mathbf{f}_M\}$ with $M>N$ (e.g., $N=16, M=32$ \cite{DeepSense}).\footnote{The proposed method in this work is general, which is thus applicable to other setups of number of antennas and codebook size.}
As such, the $m$-th beamforming vector can be expressed as
\begin{equation}
\mathbf{f}_m =
\frac{1}{\sqrt{N}}
\left[
1,
e^{j2\pi \frac{d}{\lambda}\sin(\phi_m)},
\dots,
e^{j2\pi \frac{d}{\lambda}(N-1)\sin(\phi_m)}
\right]^{\mathrm{T}},
\end{equation}
where $\phi_m$ is the azimuth angle associated with the $m$-th beam. Let \(\mathbf{w}[t] \in \mathbb{C}^{N \times 1}\) denote the BS transmit beamforming vector at time slot \(t\), which is selected from the beam codebook \(\mathcal{F}\), i.e., $\mathbf{w}[t] = \mathbf{f}_{m}$, with $\mathbf{f}_{m} \in \mathcal{F}$, where \(m \in \{1,\dots,M\}\) denotes the selected beam index at time slot \(t\). Given the BS transmit beamforming \(\mathbf{w}[t]\), the received signal at the UAV can be expressed as
\begin{equation}
y[t]= \mathbf{h}^H[t] \mathbf{w}[t] x[t] + z[t],
\end{equation}
where \(z[t] \sim \mathcal{CN}(0, \sigma^{2})\) represents the received additive white Gaussian noise (AWGN) with power \(\sigma^2\).

\subsection{Problem Formulation}
Our goal is to predict the optimal beam index for each time slot \(t\) from a predefined beamforming codebook in order to maximize the received signal power at the UAV. Specifically, given the codebook \(\mathcal{F} = \{\mathbf{f}_1, \dots, \mathbf{f}_M\}\), the optimal beam index at time slot \(t\) is given by
\begin{equation}
m^{\star}[t] =
\arg\max_{m \in \{1,\dots,M\}}
\left| \mathbf{h}^{H}[t] \mathbf{f}_{m} \right|^{2}.
\label{eq:optimal_beam_index}
\end{equation}
Accordingly, the optimal transmit beamforming vector at time $t$ is $\mathbf{w}^{\star}[t] = \mathbf{f}_{m^{\star}[t]}$. Directly solving the above beam selection problem requires an exhaustive search over all candidate beams in the codebook~\cite{YOU}, which incurs prohibitively high overhead in dynamic UAV communication scenarios. Therefore, it is crucial to develop a low-overhead beam prediction approach that can rapidly adapt to environmental dynamics.

To address this challenge, we propose a \emph{vision-based} beam prediction framework that exploits visual information captured by the BS-mounted camera, from which the channel characteristics can be inferred  from the relative locations of the transceivers and the surrounding environment. Let \(\mathbf{U}[t] \in \mathbb{R}^{W \times H \times C}\) denote the RGB image collected at time slot \(t\). The beam prediction task aims to learn a mapping function \(f_{\Theta}\), parameterized by \(\Theta\), which infers the optimal beam index based on the visual observation, i.e.,
\begin{equation}
f_{\Theta}: \mathbf{U}[t] \rightarrow m^{\star}[t].
\end{equation}

\section{Proposed BeamVLM for Beam Prediction}

\begin{figure*}[t] 
  \centering
  \includegraphics[width=0.8\textwidth]{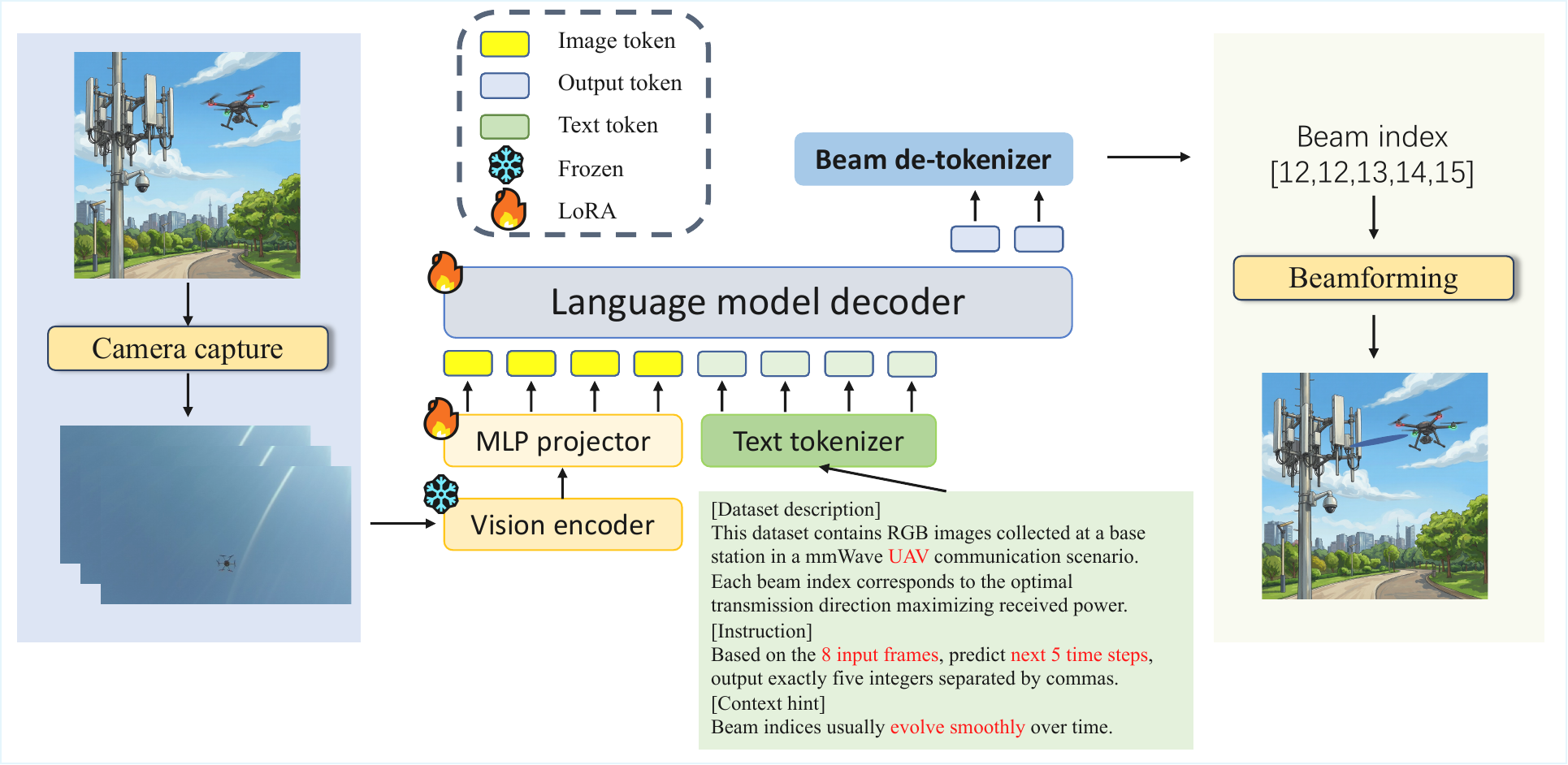} 
  \captionsetup{justification=centering}
  \caption{The framework of the proposed end-to-end generative method.}
  \label{fig:framework}
\end{figure*}

In this section, we first review the existing methods and point out their main limitations. Then, a novel BeamVLM framework is proposed, which employs \emph{multimodal} data (i.e., images and texts) to enable fast and accurate beam prediction from visual information.
\subsection{Existing Methods}
\label{sec:existing_methods}
\subsubsection{Conventional DL-based beam prediction}
The conventional DL-based beam prediction methods are usually based on discriminative deep neural networks (DNNs) such as CNN~\cite{jiang2022computer}. These approaches formulate the beam selection as a classification problem and solve it by minimizing the cross-entropy loss between the predicted beam distribution and the ground-truth labels. However, these methods lack \emph{semantic understanding} on the beam prediction task, relying heavily on fitting statistical correlations between pixel intensities and beam indices within the training distribution. As a result, when these models are deployed in unseen scenarios, the learned low-level correlations may no longer hold, resulting in performance degradation and weak generalization across other scenarios~\cite{Charan2022_UAV,jiang2022computer,camTVT}.
\subsubsection{LLM-based beam prediction}
Recently, LLMs have been introduced to the beam prediction design, such as BeamLLM~\cite{zheng2025beamllmvisionempoweredmmwavebeam} and BP-LLM~\cite{sheng2025beam}. These methods share a similar design paradigm: they utilize the LLM as a feature backbone followed by a task-specific classification head to predict beam probabilities. In terms of the input processing, BeamLLM first employs an object detector (e.g., YOLO) to extract bounding boxes from images and then feeds these coordinates into the model. For BP-LLM, it utilizes the angle-of-departure (AoD) information and historical beam indices as inputs for beam prediction.  However, the input representations in both Beam-LLM and BP-LLM methods generally suffer from information loss. This is because by focusing solely on sparse bounding boxes~\cite{zheng2025beamllmvisionempoweredmmwavebeam} or numerical values~\cite{sheng2025beam}, these methods may not fully utilize the rich environmental context, such as the height of surrounding buildings and the geometry of urban canyons, which are inherent in raw visual data. Consequently, they fail to achieve a holistic scene understanding, which is required for robust performance in complex environments. 

\subsection{Key Idea of Proposed BeamVLM}
To overcome the above limitations, we propose a novel BeamVLM framework that reformulates beam prediction from a conventional \emph{discriminative} classification problem into an \emph{end-to-end generative} task. Instead of mapping sensing features to a fixed set of beam labels, BeamVLM casts beam selection as a sequence generation problem in the language domain. By incorporating an \emph{instructional prompt} that encodes task-specific constraints and physical priors, the proposed framework guides the generation process toward structured and physically consistent beam decisions. Moreover, BeamVLM projects raw visual patches directly into the latent space
as continuous ``visual words,'' enabling the model to exploit the multimodal reasoning capability of VLM to generate the optimal beam index as discrete text tokens. As a result, BeamVLM jointly captures UAV motion, environmental geometry, and task semantics, allowing beam decisions to be inferred from holistic spatial and contextual relationships.

\subsection{Neural Network Architecture}
\label{sec:network_architecture}
The proposed BeamVLM framework is built upon the Qwen2.5-VL architecture~\cite{Qwen2.5-VL}. As illustrated in Fig.~\ref{fig:framework}, the system operates in an end-to-end generative manner. Taking a sequence of visual frames and a task-specific prompt as input, BeamVLM processes the data through aligned encoders and fuses the multimodal features with a large-scale reasoning backbone. The backbone then autoregressively generates the optimal beam indices as text tokens. In the following subsections, we introduce key modules of this architecture in detail, including the multimodal input processing, prompt embedding, visual encoders, and the backbone optimized via low-rank adaptation (LoRA). 

\subsubsection{Multimodal input processing}
\label{sec:inputs}
For our design, the overall input to the BeamVLM framework at time step $t$ is formulated as a multimodal set $\mathcal{S}[t]$, which is expressed as:
\begin{equation}
\mathcal{S}[t] = \{ \mathbf{U}[t], \mathcal{I} \},
\end{equation}
where $\mathbf{U}[t]$ and $\mathcal{I}$ denote the raw visual component and the task-specific instructional prompt, respectively. 
As illustrated in Fig.~\ref{fig:framework}, to ensure efficient feature extraction and batch processing, the raw visual data $\mathbf{U}[t]$ is first pre-processed into a standardized tensor $\mathbf{U}_{\text{norm}}$. This procedure involves resizing the image to a fixed resolution of $224 \times 224$ pixels, followed by a normalization step to stabilize the input distribution. Moreover, the instructional prompt $\mathcal{I}$ (to be detailed in Section \ref{sec:prompt_design}) is established to provide high-level semantic guidance, enabling the model to better reason about the spatial relationships within the scene. These processed inputs are subsequently fed into their respective embedding modules to generate multimodal representations.

\subsubsection{Prompt embedding module}
\label{sec:prompt}
For the input text $\mathcal{I}$, a tokenizer is utilized to transform the raw text into a discrete token sequence $\mathcal{I}_\text{t} \in \mathbb{R}^{L_\text{p} \times 1}$, where $L_\text{p}$ is the sequence length. After tokenization, these discrete tokens are mapped into continuous high-dimensional vectors, yielding text token embeddings $\mathcal{I}_\text{e} \in \mathbb{R}^{L_\text{p} \times d_\text{m}}$, with $d_\text{m}$ representing the feature dimension. This embedding process is crucial for enabling the downstream large model to comprehend the spatial information for beam prediction.  

\subsubsection{Visual encoder module}
\label{sec:Image Encoder Module}
As shown in Fig. \ref{fig:framework}, Qwen2.5-VL employs a Vision Transformer (ViT) encoder to extract high-dimensional semantic features. The normalized image $\mathbf{U}_{\text{norm}}$ is divided into a set of patches and mapped to a latent space with positional encoding. This yields patch embeddings $\mathbf{U}_\text{e} \in \mathbb{R}^{N_\text{v}\times d_\text{v}}$, where $N_\text{v}$ is the number of patches and $d_\text{v}$ denotes the dimension of the embedded features. Finally, a projection layer maps the output features to the feature dimension $d_\text{m}$, yielding the visual tokens. 

\subsubsection{Multimodal reasoning backbone}
\label{sec:backbone}
The language backbone of Qwen2.5-VL follows the architectural design of Qwen2.5. To reduce computational cost during model training, we apply the LoRA fine-tuning technique \cite{hu2021loralowrankadaptationlarge}, which is used to realize the lightweight fine-tuning by updating the query, key, and value matrices in the multi-head attention module. Given the pre-trained weight matrix $\mathbf{W}_0 \in \mathbb{R}^{d_\text{v} \times d_\text{v}}$, LoRA parameterizes the adapted weight as $\mathbf{W}=\mathbf{W}_0+\frac{\alpha}{r}\mathbf{B}\mathbf{A}$, where $\mathbf{B}\in\mathbb{R}^{d_\text{v}\times r}$ and $\mathbf{A}\in\mathbb{R}^{r\times d_\text{v}}$ are trainable low-rank matrices, and $r$ and $\alpha$ denote the rank and scaling factor, respectively. Generally, $r \ll d_\text{v}$ and the matrix $\mathbf{B}$ is initialized as a zero matrix, while the matrix $\mathbf{A}$ is initialized from a Gaussian distribution\cite{hu2021loralowrankadaptationlarge}.

Consider the multi-head attention module of the decoder block. For the $i$-th head, given the query matrix $\mathbf{Q}^i$, key matrix $\mathbf{K}^i$, and value matrix $\mathbf{V}^i$, the attention is obtained as
\begin{equation}
\mathrm{Attn}(\mathbf{Q}^i, \mathbf{K}^i, \mathbf{V}^i) = \operatorname{Softmax}\left( \frac{\mathbf{Q}_p^i(\mathbf{K}_p^i)^T}{\sqrt{d_\text{m}}} \right) \mathbf{V}^i,
\end{equation}
where $\mathbf{Q}_p^i$ and $\mathbf{K}^i_p$ are obtained by applying the rotary positional embedding, which rotates each input vector according to its position in the input sequence. 

\subsection{Output Design}
After processing the multimodal input, the backbone model generates an output token sequence in an auto-regressive manner. 
Let $\mathbf{z} = [z_1, \ldots, z_{L_{\text{token}}}]$ denote the generated sequence, where $L_{\text{token}}$ is the sequence length representing future beam indices. Specifically, at decoding step $\tau$, the model predicts the next token distribution $P_\Theta (z_\tau \mid \mathcal{S}[t], \mathbf{z}_{<\tau})$ conditioned on the multimodal input $\mathcal{S}[t]$ and the sequence of previously generated tokens $\mathbf{z}_{<\tau}= \{z_1, z_2, \dots, z_{\tau-1}\}$, where $\Theta$ denotes the trainable parameters of BeamVLM.

\textbf{Training stage}: During training, we employ a \textit{teacher forcing} technique to minimize the cross-entropy loss between the predicted distribution and the ground-truth sequence $\mathbf{z}^\star$:
\begin{equation}
\mathcal{L}(\Theta) = - \sum_{\tau=1}^{L_{\text{token}}} \log P_\Theta \left( z_\tau^\star \mid \mathcal{S}[t], \mathbf{z}_{<\tau}^\star \right).
\end{equation}

\textbf{Inference stage}: In the inference stage, the generated tokens are mapped back to physical beam indices via a \textit{de-tokenization} process, ensuring that the linguistic outputs are projected onto the discrete beam codebook $\mathcal{F}$.

\subsection{Prompt Engineering}
\label{sec:prompt_design}
To effectively bridge the gap between visual perception and physical-layer decision-making, the design of instructional prompt $\mathcal{I}$ is of paramount importance. Instead of treating the VLM as a generic image captioner, we construct a \emph{structured} prompt that explicitly injects domain knowledge and imposes strict output constraints. As illustrated in Fig.~\ref{fig:framework}, the prompt $\mathcal{I}$ is composed of three concrete information blocks, which does not need to update in real-time:

\subsubsection{Dataset definition}
The prompt explicitly specifies the dataset attributes, including the number of consecutive input frames (e.g., 8 frames captured from the BS at $t=-7,\ldots,0$) as well as the codebook size (e.g., $|\mathcal{F}|=32$), which constrains the beam search space.

\subsubsection{Task instruction and constraints}
The core instruction directs the model to analyze the flight trajectory from the input frames and predict the beam indices for the subsequent 5 time slots ($t=1 \dots 5$). 
To ensure that the output is directly parsable for beam alignment, we impose a strict formatting constraint on the model output, enforcing it to produce exactly five comma-separated integers corresponding to the predicted beam indices (as illustrated in Fig.~\ref{fig:framework}).

\subsubsection{Context hint}
To enhance reasoning accuracy, we inject specific physical priors into the prompt. In particular, we inform that ``beam indices usually evolve smoothly over time'' and instruct the model to utilize this \emph{temporal stationarity}. This hint  encourages the model to avoid abrupt beam jumps that seldom occur in practice and generate temporally consistent beam sequences based on the UAV’s continuous flight path dictated by physical inertia.

\begin{table}[t]
\centering
\caption{Parameter settings}
\label{tab:parameters}
\begin{tabular}{l r l r}
\toprule
\textbf{Parameter} & \textbf{Value} & \textbf{Parameter} & \textbf{Value} \\
\midrule
Codebook size ($M$) & 32 & Batch size & 16 \\
Optimizer & AdamW & Learning rate & $1\times 10^{-5}$ \\
Weight decay & $1\times 10^{-2}$ & BS antennas ($N$) & 16 \\
\bottomrule
\end{tabular}
\vspace{-3mm}
\end{table}

\begin{figure*}[t] 
    \centering
    \begin{minipage}[b]{0.32\textwidth}
        \centering
        \includegraphics[width=\textwidth]{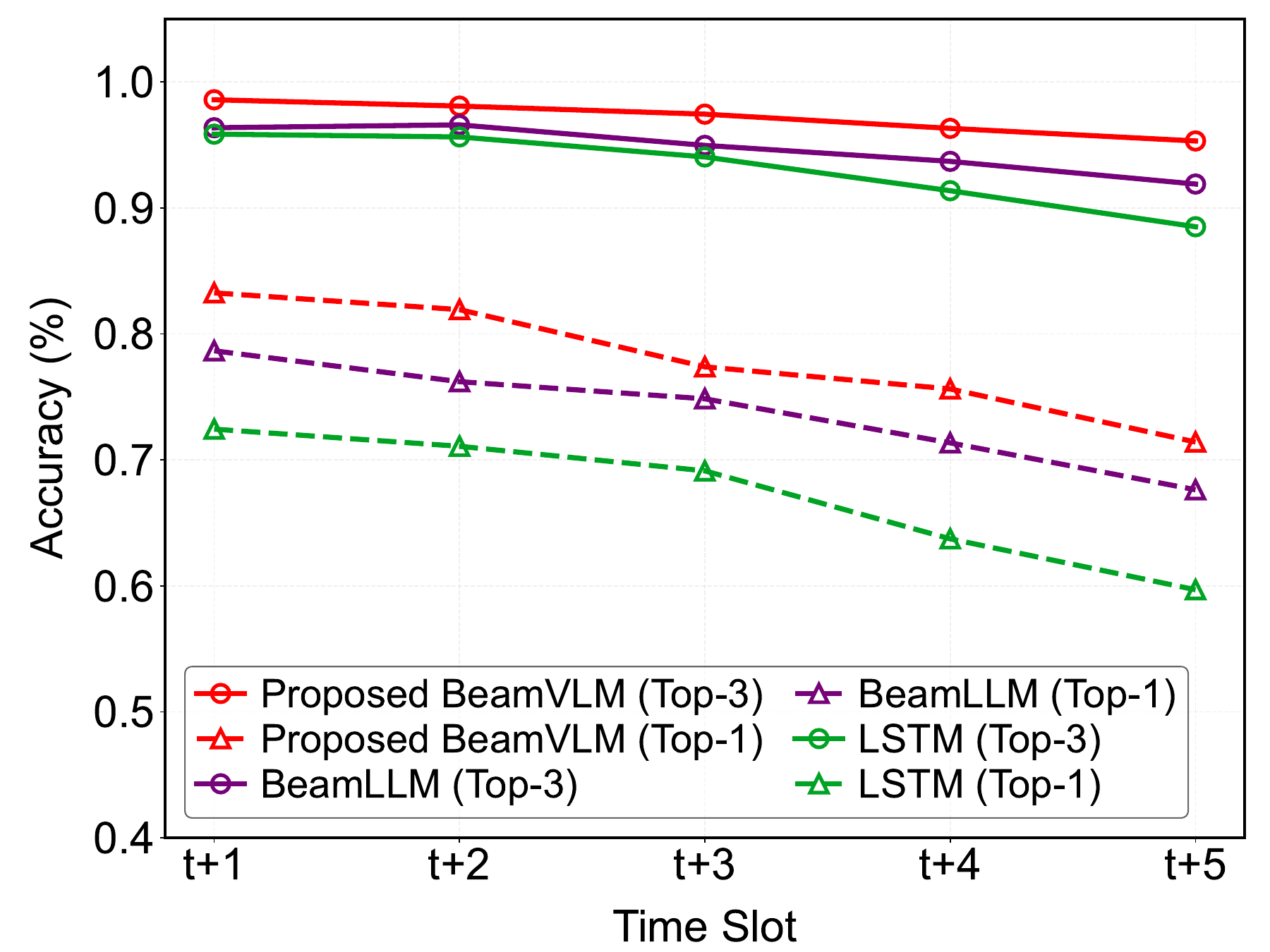}
        \caption{Top-$K$ accuracy over future time slots in UAV scenarios.}
        \label{fig:uav_performance}
    \end{minipage}
    \hfill 
    \begin{minipage}[b]{0.32\textwidth}
        \centering
        \includegraphics[width=\textwidth]{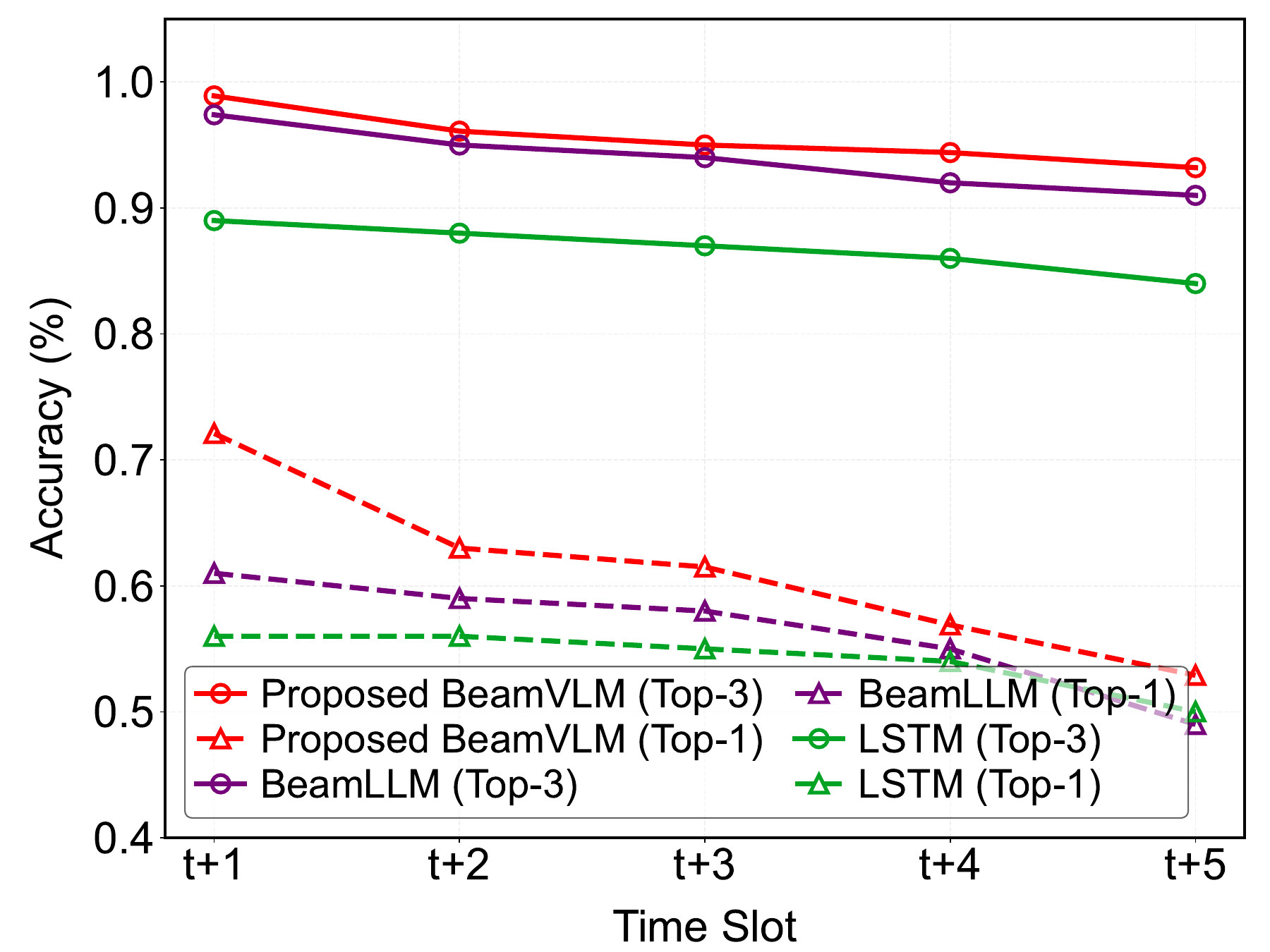}
        \caption{Top-$K$ accuracy over future time slots in V2I scenarios.}
        \label{fig:v2i_performance}
    \end{minipage}
    \hfill 
    \begin{minipage}[b]{0.32\textwidth}
        \centering
        \includegraphics[width=\textwidth]{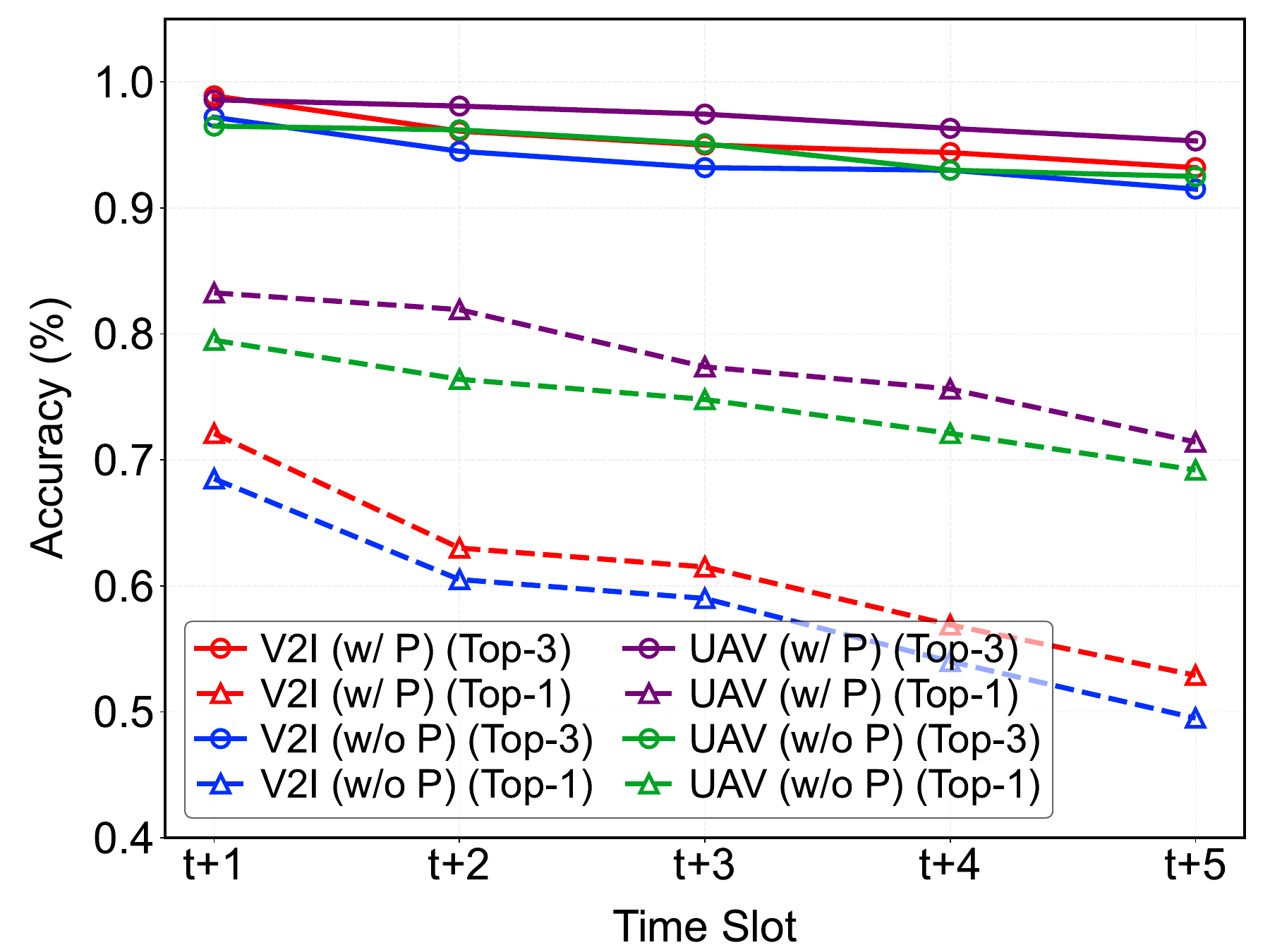} 
        \caption{Ablation study of prompt in UAV and V2I scenarios.}
        \label{fig:ablation_study}
    \end{minipage}
\end{figure*}

\section{Performance Evaluation}
\subsection{Experimental Setup}\label{experiment}
Dataset processing: For performance evaluation, we conduct extensive experiments using a real-world mmWave ground-to-UAV dataset, DeepSense 6G Scenario 23~\cite{DeepSense}. For evaluation, 70\% of the samples from a single scene are used for fine-tuning, and the remaining 30\% are used for testing. Each data sequence is decomposed into samples using a sliding window of size 13, consisting of 8 input frames and 5 future prediction steps.

Parameter settings: The proposed BeamVLM is implemented based on the pre-trained Qwen2.5-VL model with 3B parameters and fine-tuned using LoRA. Following standard practices in parameter-efficient adaptation~\cite{hu2021loralowrankadaptationlarge}, the LoRA rank is set to $r = 8$ with a scaling factor of $\alpha = 16$. Other training hyperparameters are summarized in Table~\ref{tab:parameters}. All experiments are implemented in PyTorch and conducted on two NVIDIA RTX 4090 GPUs.  

Baselines: To evaluate the performance of the proposed BeamVLM, several representative beam prediction approaches are selected as baselines, including RNN~\cite{jiang2022computer}, LSTM, and BeamLLM~\cite{zheng2025beamllmvisionempoweredmmwavebeam}.
All methods are trained and tested on the same dataset splits to ensure a fair comparison.

Evaluation metrics:
To evaluate the beam prediction performance, we employ Top-$K$ accuracy, which measures the percentage of test samples where the ground-truth beam index is captured within the $K$ most probable predicted candidates.

\begin{table}[t]
\centering
\caption{Model complexity and inference efficiency.}
\label{tab:complexity_time}
\renewcommand{\arraystretch}{1.05}
\setlength{\tabcolsep}{8pt}
\begin{tabular}{l l l r}
\toprule
\textbf{Model} & \textbf{Total Para.}& \textbf{Trainable Para.}& \textbf{Runtime (s)}\\
\midrule
LSTM           & 104.4K & 104.4K & $7.2\times 10^{-5}$ \\
BeamLLM        & 178.3M & 53.9M  & $2.3\times 10^{-3}$ \\
BeamVLM (Ours) & 3.1B   & 42.2M  & $9.5\times 10^{-2}$ \\
\bottomrule
\end{tabular}
\vspace{-2mm}

\end{table}

\subsection{Numerical Results}
\subsubsection{UAV scenario performance}
Fig.~\ref{fig:uav_performance} illustrates the beam prediction accuracy for the UAV scenario. It is observed that the proposed BeamVLM demonstrates superior performance in Top-1 accuracy, achieving 83.3\% at $t+1$ and 71.4\% at $t+5$. This represents a significant improvement over the LSTM baseline, achieving a performance gain of 10.8\%. The robustness of BeamVLM is further evidenced under the Top-3 accuracy, which remains remarkably high throughout the entire prediction horizon, while baselines such as BeamLLM and LSTM experience pronounced degradation, falling to 91.9\% and 88.5\% at $t+5$, respectively. In terms of average runtime in reference shown in Table~\ref{tab:complexity_time}, due to more involved computation, BeamVLM requires a longer time than baselines.

\subsubsection{Generalization in V2I scenarios}
To evaluate generalization, we retrained BeamVLM on the Scenario 8 dataset~\cite{DeepSense}. The textual prompt is adjusted to reflect V2I-specific semantics. As shown in Fig.~\ref{fig:v2i_performance}, the proposed BeamVLM consistently achieves the highest Top-1 accuracy, reaching 72.1\% at $t+1$, which outperforms BeamLLM, LSTM, and RNN by 11.1\%, 16.1\%, and 26.0\%, respectively. 
At $t+5$, it sustains 52.9\% accuracy, maintaining a 4\% lead over the second-best model. 
For Top-3 accuracy, BeamVLM remains above 93.0\% throughout all horizons, whereas baselines like BeamLLM and LSTM drop to 91.0\% and 84.0\% at $t+5$. 
These results confirm that proposed multimodal reasoning enables robust beam prediction even in dynamic V2I environments where conventional sequence models degrade rapidly.

\subsubsection{Ablation on prompt guidance}
To evaluate the importance of prompt guidance, we compare BeamVLM with a variant, \emph{BeamVLM w/o Prompt}, where textual instructions are removed. As shown in Fig.~\ref{fig:ablation_study}, removing the prompt results in an accuracy drop across both UAV and V2I scenarios. Specifically, the initial Top-1 accuracy reduces by 3.76\% and 3.60\% for the UAV and V2I scenarios, respectively. This gap confirms that while visual patches capture geometric context, textual prompts provide essential semantic guidance for precise beam inference. Notably, even without prompts, the model still outperforms BeamLLM and LSTM, validating the robustness of our generative vision-language architecture. These results reveal that the textual guidance is key to the superior performance of BeamVLM. 

\section{Conclusions}
In this work, we proposed a novel VLM-based framework, called BeamVLM, for mmWave beam prediction in LAE. By leveraging the pretrained multimodal knowledge of large VLMs and task-oriented prompt design, the proposed BeamVLM reformulates beam prediction as a generative reasoning task, enabling interpretable and end-to-end prediction without handcrafted output heads. Experimental evaluations demonstrated that BeamVLM achieves superior accuracy compared with various baselines. In addition, numerical results also showed that BeamVLM exhibits excellent generalization performance in the V2I scenario.

\bibliographystyle{IEEEtran}
\bibliography{refs}

\end{document}